\documentclass[twocolumn,   preprintnumbers,amsmath,amssymb]{revtex4}
\usepackage{graphicx}% Include figure files
\usepackage{dcolumn}% Align table columns on decimal point
\usepackage{bm}% bold math
\begin{document}
\title{Layer-by-layer assembly of colloidal particles deposited onto the polymer-grafted elastic substrate}
\author{Kang Chen and Yu-qiang Ma}

\altaffiliation[ ]{Author to whom correspondence should be
addressed. Electronic mail: myqiang@nju.edu.cn.}
\address{National Laboratory of Solid State Microstructures,
Nanjing University, Nanjing 210093, China}

\begin{abstract}
We demonstrate a novel route  of spatially organizing the colloid
arrangements on the polymer-grafted substrate by use of
self-consistent field and density functional theories. We find
that grafting of polymers onto a substrate can effectively control
spatial dispersions of deposited colloids as a result of the
balance between colloidal settling force and entropically elastic
force of brushes, and colloids can form unexpected ordered
structures on a grafting substrate. The depositing process of
colloidal particles onto the elastic ``soft" substrate includes
two steps: brush-mediated one-dimensional arrangement of colloidal
crystals and controlled layer-by-layer growth driven entropically
by non-adsorbing polymer solvent with  increasing the particles.
The result indicates a possibility for the production of highly
ordered and defect-free structures by simply using the grafted
substrate instead of periodically patterned templates, under
appropriate selection of colloidal size, effective depositing
potential, and brush coverage density.
\end{abstract}
\maketitle

Colloidal particles can self-assemble into a rich variety  of
highly ordered structures\cite{whitesides} on periodically
patterned substrate \cite{Wiltzius},  block copolymer scaffolds
\cite{lopes}, vesicle surfaces with the opposite charge
\cite{ramos}, and at brush/air interfaces \cite{kim,kim2} and
liquid/liquid \cite{xxx} or water/air \cite{xxxx} interfaces. The
major challenges in this field are how to assemble monodispersed
 colloids into highly ordered structures with
well-controlled sizes and  shapes, and  how to achieve
layer-by-layer growth of ordered structures. Sedimentation  is  a
simplest approach for   colloidal crystallization, however,
usually leads to uniform or simple close-packed arrays of
colloidal particles on smooth substrates. Effective control over
interaction and arrangement of colloids\cite{kim2,991} is possible
by using densely polymer-grafted substrates\cite{99}, since the
entropically elastic energy of  brushes is comparable to thermal
energy, and self-assembly depends critically on thermal energy
\cite{ramos}. A balance between depositing force of colloids and
entropic force of brushes probably leads to the formation of
ordered colloidal crystal structures, rather than highly
disordered aggregates.

The grafting of polymers to surfaces is a simple and useful
approach to stabilize colloids against aggregation and adsorption.
It was experimentally reported \cite{sbo} that the thickness   of
brushes  can  be adjusted between  several nm and 1 $\mu m$. The
grafted polymer always exerts a repulsive entropic force on
incoming particles, and past works focused on the interaction
between polymer brushes and individual incoming particles and how
to prevent the adsorption of colloids such as proteins onto
surfaces under various grafting density, chain length, and
interactions between chains and surface\cite{99}. To the best of
our knowledge, there have been no systematic theoretical studies
into the self-assembling structures of  colloidal particles when
deposited onto the grafting substrate. Here, we undertake the
first theoretical study of deposition of colloidal particles from
polymer solvent onto the grafted substrate, and address an
important issue about how to realize the spatial organizing
dispersions of colloidal particles.

We consider a mixture of $n_{\beta}$   solvent chains and $n_p$
colloidal particles with the radius $R$, in contact with a
substrate grafted with $n_{\alpha}$ polymer chains. The  substrate
 is horizontally placed in $xy$-plane which is
positioned at $z=0$,  and     all polymer chains and colloids are
allowed in the region $0\leq z\leq Z_{max}$.   $Z_{max}$ is large
enough to avoid the influence of brushes and particles, namely the
solvent chains attain their bulk property for large values of $z$
\cite{13}. It is shown that  aggregates on the deformed polymer
brush are anisotropic,  and extend along only one direction for
minimizing the entropy loss of brush chain
conformations\cite{kim2}. Here, the existence of ``settling" force
helps to form horizontally oriented cylinders. Thus we assume
translational invariance along y for the sake of simplicity, and
the calculation can be reduced to the xz plane. The volume of the
system $V$ is $L_x\times Z_{max}$, where $L_x$ is the lateral
length of the surfaces along the $x$ axes. The grafting density is
defined as $\sigma =n_{\alpha}/L_x$. All polymer chains are of the
same polymerization index $N$ and flexible with the same
statistical length $a$, and incompressible with a segment volume
$\rho _0^{-1}$. The probability distribution for  molecular
conformations of a Gaussian chain $\alpha$  is assumed to take the
Wiener form $
\textit{P}[\mathbf{r}_{\alpha}(s)]\varpropto\exp[-\frac{3}{2a^{2}}\int_{0}^{N}ds|\frac{d\mathbf{r}_{\alpha}(s)}{ds}|^{2}]
$, where ${\bf r}_\alpha(s)$ denotes the position of segment s on
chain $\alpha$. Recently, Balazs and co-authors \cite{5} have
successfully combined a self-consistent field
 (SCF) theory with a  density
functional theory (DFT) to study mixtures of diblock copolymer and
nanoparticles. Here, we use  the grand canonical form of SCF
theory which has been proven to be powerful in calculating
equilibrium morphologies in polymeric system \cite{13,5,14,15}, to
deal with polymer solvent and brushes, while  particles are
treated by DFT \cite{Den,sss} to determine their favorite
distributions. The grand canonical partition function \cite{13}
for the system can be written as $
Z_\mu=\sum_{n_\beta =0}^{\infty} e^{n_\beta N \mu}%
Z_{n_\alpha , n_\beta, n_p}$, where $\mu$ is the chemical
potential per solvent segment  and $Z_{n_\alpha,n_\beta,n_p}$ is
the canonical partition function for $n_\alpha$ grafted chains,
$n_\beta$ solvent polymers, and $n_p$ particles:

\begin{eqnarray}
Z_{n_\alpha, n_\beta, n_p} &=&\frac{1}{n_\beta !} \frac{1}{n_p !}%
\int\prod_{\alpha =1}^{n_\alpha}\textit{D}\mathbf{r}_{\alpha}(s)%
\textit{P}[\mathbf{r}_\alpha(s)]\prod_{\beta%
=1}^{n_\beta}\textit{D}\mathbf{r}_{\beta}(s)\nonumber\\
&&\textit{P}[\mathbf{r}_\beta(s)]\prod_{p=1}^{n_p}\textit{d}{\bf R_p}%
\delta[1-\widehat{\varphi}-\widehat{\varphi}_s
-\widehat{\varphi}_p]\nonumber\\
&&\exp[-{\frac{\nu}{k_B
T}}]\prod_{\alpha=1}^{n_\alpha}\delta(\mathbf{r}_\alpha(0)-\mathbf{r}^\alpha)\;,
\end{eqnarray}
where  the first $\delta$ function enforces incompressibility and
the second $\delta$ function determines the  position of the
uniformly anchored chain end. $k_B$ is Boltzmann's constant, T is
the temperature, and $\nu$ is the interaction energy. ${\bf R_p}$
is the position of the center of the $p$th particle.
  The operators $\widehat{\varphi}$,
$\widehat{\varphi}_s$, and $\widehat{\varphi}_p$ represent the
local concentrations  of grafted chains, solvent, and colloidal
particles, respectively. The SCF theory gives the free energy $F$
\begin{eqnarray}
\frac{N F}{\rho_0 k_B TV} &=&-\phi (\frac{1}{n_\alpha} \sum_{\alpha =1}^{n_\alpha} \ln  Q_{\alpha})%
-\frac{N}{\rho_0 V} e^{N \mu}  Q_S-\frac{N\phi_p}{\rho_0 \pi R^2}
\nonumber\\
&&\ln
\frac{ Q_p}{V\phi_p}+\frac{N  \nu}{\rho_0 k_B T V}-\frac{1}{V} \int \textit{d}{\bf r}%
[\xi(1-\varphi-\varphi_s \nonumber\\
&&-\varphi_p)
+w \varphi+w_s \varphi_s +w_p \rho_p -\rho_p%
\Psi(\overline{\varphi}_p)]\;\;,
\end{eqnarray}
where   $ \varphi$, $\varphi_s$, and $\varphi_p$ are  the local
volume fractions of brushes, solvent chains, and particles, and
 the  overall volume fractions of brushes and particles are given by $\phi$ and $\phi_p$. $\rho_p$ stands for the
particle center distribution, and the local particle
volume fraction is then given by $\varphi_p({\bf r}) =\frac{\rho_0}{N} \int_{|{\bf r'}|<R} \textit{d}{\bf r'}%
\rho_p({\bf r}+{\bf r'})$ \cite{5}.
$ Q_{\alpha}  =\int \textit{d}{\bf r}%
q_\alpha({\bf r},s)q_{\alpha}^{\dag}({\bf r},s) $
 represents the single chain partition
function of brushes  subject to the field $w$, and $ Q_s =\int
\textit{d}{\bf r} q_s({\bf r},s)q_s({\bf r},1-s)$ and $ Q_p =\int
\textit{d}{\bf r} \exp[-w_p({\bf r})]$ are the partition functions
for solvent and particles under fields $w_s$ and $w_p$,
respectively. The end-segment distribution functions
$q_{i}(\mathbf{r},s)$ and $q_{i}^{\dag}(\mathbf{r},s)$ represent
the probability of finding the $s^{th}$ segment  at position
$\bold r$ respectively from two distinct ends of chains. $q_{i}$
satisfies a modified diffusion equation $ \frac{\partial
q_{i}}{\partial s}=\frac{a^2}6\nabla ^2q_{i}-w_{i}({\bf r}) q_{i}
$, and $ q_i^{\dag}$ meets the same diffusion equation but with
the right-hand side multiplied by $-1$.   The last term in Eq.(2)
is DFT term \cite{Den} accounting for the steric interaction
between particles, and the excess free energy
$\Psi(\overline{\varphi}_p)$ per particle is from the
Carnahan-Starling function \cite{sss} with the weighted particle
density, $ \overline{\varphi}_p({\bf r})$ \cite{5}. The
interaction energy $\nu$ is given by $
 {N   \nu}/{\rho_0 k_B T}= \int \textit{d} {\bf r} [\chi_{bs} N \varphi \varphi_{s}%
+\chi_{bp} N \varphi \varphi_p +\chi_{sp} N \varphi_s \varphi_p+
 g_e z \varphi_p],$ where the $\chi$'s are the Flory-Huggins
interaction parameters between the different chemical species. We
fix $N=100$, $\chi_{bs}N=0$, and $\chi_{bp}N=\chi_{sp}N=12.0$,
since we assume that all the polymers have the same chemical
nature, while the particles are insoluble to polymers\cite{sma}.
In addition,   a depositing force is applied normal to the
substrate, and $g_e >0$ is the strength of settling field acting
on particles. Here, we choose $g_e$ within the range $0.1 \sim
0.6$ which, on  one hand, can ensure that the size of
particles($2R=0.5R_0$) is comparable to the sedimentation length
($\propto \frac{N}{\pi R^2 g_e }$ \cite{sda}), $0.8 R_0\sim5.0
R_0$, where $R_0\equiv aN^{1/2}$ characterizes the natural size of
polymer.  And on the other hand, the competition between
depositing energy of colloids and elastic entropy of deformed
brushes can significantly appear within this range of $g_e$. The
volume fraction $\phi_p$ of particles  changes from $0.03$ to
$0.15$ for ensuring the possibility of colloidal crystallization
on the brush surface  and   layer-by-layer growth of colloids. For
sub-micrometer particles, the depositing force may be introduced
either by gravity or by centrifuge\cite{sda,sssso}, whereas for
nano-sized colloids, the sedimentation may be controlled via
electrophoretic techniques \cite{sssso,sssso1,ssb}. We take
$\sigma=0.25$ so that particles hardly embed into the dense brush.
A periodical boundary condition for the $x$-direction is applied,
while for the $z$-direction, the region of $z<0$ is forbidden and
the region  $z>Z_{max}$ is treated as a bath of solvent. $Z_{max}$
is fixed to be $60a$, and the lateral length $L_x$ is selected to
minimize the free energy of system \cite{wang}. All the sizes are
in units of $a$. In SCF theory, the fields and densities  are
determined by locating saddle points in the free energy subject to
the incompressibility: $\varphi({\bf r})+\varphi_s({\bf
r})+\varphi_p({\bf r})=1$. The resulting SCF equations are solved
by the combinatorial screening algorithm of Drolet and Fredrickson
\cite{15}.
\begin{figure}
\includegraphics[width=8.4cm]{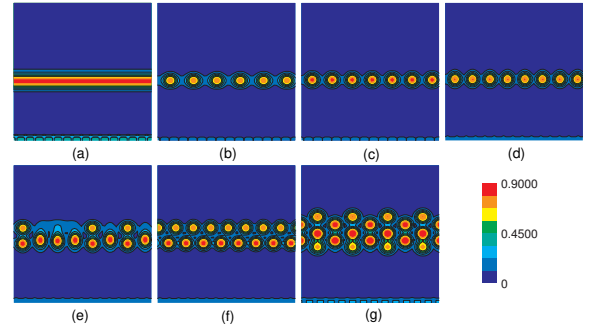}
\caption{  Particles density distributions in $x$-$z$
cross-sections   under different particle volume fractions. (a)
$\phi_p=0.03$, (b) $\phi_p=0.04$, (c) $\phi_p=0.05$, (d)
$\phi_p=0.06$,  (e) $\phi_p=0.092$, (f) $\phi_p=0.1$, and (g)
$\phi_p=0.145$. The color scale bar shows the local density values
of particles in Figs. 1(a)-1(g). }
\end{figure}

We first examine the effects of colloid volume fraction $\phi_p$
on colloidal dispersions at the top of brushes. Figure 1 shows the
morphology changes of dispersed particles with increasing $\phi_p$
for the parameter $g_e=0.5$. We find that at low particle volume
fraction (Fig.1(a)), the dispersion of particles forms a laterally
uniform layer on the top of brushes. This means that the
deposition of particles does not deform brushes to in turn react
with the dispersion of particles, and particles only spread over
the grafting surface, which is similar to that of particles
depositing onto the smooth substrate. As more colloidal particles
are deposited, the grafted polymer is compressed under depositing
potential, and it responds with a restoring force which further
drives particles to self-assemble into certain structures for
counteracting deposition of particles, in the requirement of
minimizing combinational contribution of colloidal depositing
energy and brush entropy. Therefore, colloidal particles assemble
into colloidal crystals under
 sedimentation, and morphology of regularly separated cylinder
structures emerges. The number of cylinders will increase with the
 volume fraction $\phi_p$ (Figs. 1(b)-1(d)). However,
further increase of  $\phi_p$ leads to the formation of second
layer of cylinders(Figs. 1(e) and 1(f)) piled on the first layer
created by brushes, and even forms the third layer of cylinder
structures(Fig.1(g)).
\begin{figure}
\includegraphics[width=7.90cm]{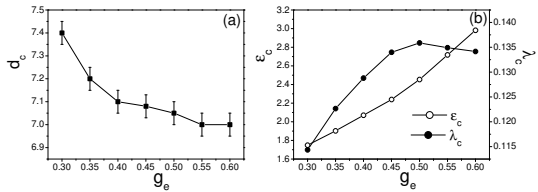}
\caption{ (a) The minimum spacing $d_c$ between cylinders  vs
$g_e$ with error bars. (b) The critical   correlation factor
$\lambda_c$ and penetration depth $\varepsilon_c$ as a function of
$g_e$.}
\end{figure}

The  first-layered colloidal structure is formed, due to the
competition between colloidal sedimentation and brush entropy
effects. For a fixed $g_e$, the deposited number of cylinders of
radius R increases with   increasing $\phi_p$ when the variation
of depositing potential per cylinder $ g_e \varepsilon \pi R^2 k_B
T/N \geq \Delta F_b $. Here, $\Delta F_b $ represents a brush
stretching energy penalty of a local single cylinder with the
penetration depth $\varepsilon$ into the brush \cite{kang}, which
is given by $\Delta F_b = \frac{1}{2}\eta \varepsilon^2\pi$
\cite{Fred}.  For melt brush, the shear modulus
$\eta=\eta_0=3\sigma^2 k_BT$. We introduce a modified factor
$\lambda$ (i.e., $\eta=\lambda\eta_0$)
 to account for the correlation effect due to the interplay between multi-cylinders mediated by
brushes. Therefore,
 the variation of $\lambda$ will mainly depend on the  distance $d$
between cylinders. However, we should point out that here,  the
density profiles of free ends of brushes are diffusive due to the
existence of   solvent chains above  brushes. This will lead to
the great decrease of shear modulus compared to that of melt
brushes. We have $ \lambda \leq \frac{2 g_e
R^2}{3N\sigma^2\varepsilon},$ and the equality is valid at the
critical  $\phi_p$ where the minimum spacing between cylinders is
reached, and larger $\phi_p$ will lead to second-layer aggregation
of particles. From the SCFT/DFT calculations, we determine the
minimum spacing $d_c$ (Fig. 2(a)) and the critical penetration
depth $\varepsilon_c$ of cylinders(Fig. 2(b)), and find that $d_c$
decreases, whereas  $\varepsilon_c$   increases with $g_e$. This
clearly shows that with increasing $g_e$, depositing potential of
particles is  balanced by further deformation of brushes. Figure
2(b) also gives the critical modified factor $\lambda_c$ as a
function of $g_e$, showing that at critical $\phi_p$, the
effective shear modulus first increases with increasing $g_e$ and
then slightly decreases at large $g_e$. For small $g_e$, the shear
modulus increases with $g_e$, indicating that the correlation
between cylinders is enhanced due to the decrease of   $d_c$. In
contrast, for large $g_e$, the minimum spacing between cylinders
keeps almost unchanged, but the embedding depth $\varepsilon_c$
becomes larger than the cylinder radius so that the grafted chains
can cross the narrow gap between cylinders and fill the upper
space. Thus, the grafted chains are slightly released, which
accounts for small decrease of $\lambda_c$ at large $g_e$($>0.5$).
\begin{figure}
\includegraphics[width=7.90cm]{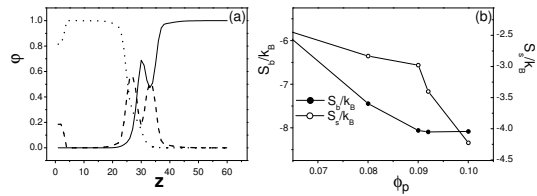}
\caption{(a) Lateral statistical concentration morphology of
brush(dotted curve), free chains(solid curve), and
particles(dashed curve) for $\phi_p=0.1$. (b) Entropies of brushes
($  S_b$) and solvent ($ S_s$) vs $\phi_p$. }
\end{figure}

The formed first-layer colloidal crystal may serve as a template
for   next-layer structural formation, and thus provides a
possible route to the fabrication of multi-layer microstructures.
We find from Fig. 1(f) that the second layer is well arranged,
based on the already deposited layer. However, the morphology
selection of second-layer colloidal dispersions will be affected
by the entropic effect of solvent polymers.  By calculating the
$z$-direction averaged density profiles of brushes, free chains,
and particles for the case of Fig.1(f), Fig. 3(a) clearly shows
that more solvent chains will fill the space between the first
layer and the second layer of cylinders. Therefore, the
second-layer cylinder structure is selected  to
 alternately arrange with the first
layer for increasing the configurational entropy of confined
solvent chains. In fact, the alternating arrangement of cylinders
in Fig. 1(g) further supports our viewpoint on controlled
layer-by-layer growth driven entropically by  polymer solvent.
Figure 3(b) gives the entropies of brush and solvent as a function
of $\phi_p$. We find that when $\phi_p$ takes the range $0.09 \sim
0.10$ corresponding to the forming process of second-layer
structure, the brush entropy retains almost unchanged, while the
solvent entropy sharply declines, meaning that the second-layer
colloidal assembly is out of brush effects, instead the solvent
entropy dominates the final equilibrium dispersion of the
second-layer particles.
\begin{figure}
\includegraphics[width=8.5cm]{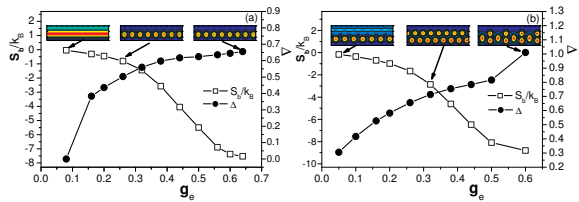}
\caption{ The brush entropy $  S_b$ and   the modulated
fluctuation
 $\Delta$ of brush height vs $g_e$.
(a)$\phi_p=0.06$. (b) $\phi_p=0.1$.}
\end{figure}

 Finally, Fig.4 shows the entropy of brushes and the
root-mean-square fluctuation $\Delta$ of  statistical brush height
$h$ \cite{kang} as a function of $g_e$  for $\phi_p=0.06$ and
$0.1$. We see that the brush entropy decreases with an increase of
$g_e$, but the height fluctuation due to brush deformation
increases with $g_e$. Correspondingly, the particle distributions
in the inset of Fig.4 signify the formation of  cylinder
structures with varying $g_e$. Figure 4(a) shows that for small
$g_e$,  the deposition of particles did not deform  brushes which
therefore do not react with  the dispersion of particles. On the
other hand, the relatively large range of the parameter $g_e$ can
stabilize the one-layer cylinder structures due to strongly
entropic restoring forces of brushes. In contrast, Fig. 4(b) shows
that a small range of $g_e$ may retain the two-layer cylinder
structures. When $g_e$ is small, there are not enough particles
deposited onto the top of brushes, leading to one-layer structure.
As $g_e$ is relatively large, the two-layer structure is
destroyed, instead the alternating structure of one- and bi-layer
cylinders appears, because the entropy restoring force of solvent
chains is weaker than that of brushes, and may not completely
offset the settling energy of particles if the second layer is
formed. It is actually interesting that the brushes have large
entropic restoring forces which easily stabilize colloidal
dispersions. For example, depending on the colloidal weight and
volume fraction, brushes can adjust the number of cylinders
formed, in contrast to colloidal crystallization  in non-adsorbing
polymer solvent.

In summary, we have demonstrated that under suitable density and
depositing force of particles, colloidal particles  can be sorted
into  alternating  arrays of cylinders by use of grafted
substrates.  The colloidal dispersions are dominated by the
requirement of minimizing combinational contribution of depositing
potential of particles and entropic restoring force of the
deformed brushes. With an increase of colloidal additions,
controlled layer-by-layer growth is driven by entropic effects of
solvent chains. The advantage of the present approach is that
control over arrangement of colloids did not rely on other
patterned \cite{Wiltzius} and phase-separated copolymer
\cite{lopes} templates but was achievable via a polymer-grafted
substrate which is easily manufactured. The approach that under
sedimentation, polymer entropic restoring force drives ordered
structure formation, will offer a simple and powerful alternative
for producing 2D and even 3D structures, and may open up an
unexplored route for engineering highly ordered structures from
colloidal building blocks.

This work was supported by  the National Natural Science
Foundation of China  under Grant Nos. 10334020,  10021001, and
20490220.


\begin{references}

\bibitem{whitesides}G. W. Whitesides and M. Boncheva, \textit{Proc. Natl. Acad. Sci.}
\textbf{99}, 4769 (2002); J. Zhu et al., \textit{Nature}
\textbf{387}, 883(1997).
\bibitem{Wiltzius}A. V. Blaaderen, R. Ruel, and P. Wiltzius, \textit{Nature}
\textbf{385}, 321(1997); K. Lin et al., \textit{Phys. Rev. Lett.}
\textbf{85}, 1770(2000); Y. Yin et al., \textit{J. Am. Chem. Soc.}
\textbf{123}, 8718(2001).
\bibitem{lopes}W. A. Lopes and H. M. Jaeger, \textit{Nature}
\textbf{414}, 735(2001); M. J. Misner et al.,  \textit{Adv.
Mater.} \textbf{15}, 221(2003); M. R. Bockstaller et al.,
\textit{J. Am. Chem. Soc.} \textbf{125}, 5276 (2003).
\bibitem{ramos}L. Ramos et al,   \textit{Science}
\textbf{286}, 2325(1999).
\bibitem{kim}Z. Liu  et al., \textit{Nano. Lett.} \textbf{2},
 219(2002).
\bibitem{kim2} J. U. Kim and B. O'Shaughnessy, \textit{Phys. Rev. Lett.} \textbf{89}, 238301 (2002).

\bibitem{xxx} Y. Lin et al.,  \textit{Science}
\textbf{299}, 226(2003).
\bibitem{xxxx} H. Wickman and J.N. Korley,
\textit{Nature} \textbf{393}, 445(1998).
\bibitem{991}K. Chen and Y. Ma, \textit{J. Phys. Chem. B}
\textbf{109}, 17617(2005); C. Ren and Y. Ma, \textit{J. Am. Chem.
Soc.} \textbf{128},   2733(2006).
\bibitem{99}S. T. Milner,  \textit{Science}
\textbf{251}, 905(1991);  J. Satulovsky, M. A. Carignano, and I.
Szleifer, \textit{Proc. Natl. Acad. Sci.} \textbf{97}, 9037
(2000).
\bibitem{sbo}M. Biesalski  and J. Ruhe,  \textit{Macromolecules} \textbf{32},
2309(1999);  J. Habicht, M. Schmidt, J.  Ruhe, and D. Johannsmann,
\textit{Langmuir} \textbf{15}, 2460(1999).
\bibitem{13} P. G. Ferreira, A. Ajdari and L. Leibler,
\textit{Macromolecules} \textbf{31}, 3994 (1998).
\bibitem{5} R. B. Thompson et al., \textit{Science} \textbf{292}, 2469 (2001);
\textit{Macromolecules} \textbf{35}, 1060 (2002).
\bibitem{14}M. W. Matsen and M. Schick, \textit{Phys. Rev. Lett.} \textbf{72}%
, 2660(1994); D. Petera and M. Muthukumar, \textit{J. Chem. Phys.}
\textbf{109}, 5101(1998); F. Schmid, \textit{J. Phys.:Condens.
Matter} \textbf{10}, 8105 (1998); M. W. Matsen and F. S. Bates,
\textit{J. Chem. Phys.} \textbf{106}, 2436(1997); P. Maniadis et
al., \textit{Phys. Rev. E} \textbf{69}, 031801(2004);  M. Muller,
\textit{Phys. Rev. E} \textbf{65}, 030802(2002); T. Geisinger,  M.
Muller, and  K. Binder, \textit{J. Chem. Phys.} \textbf{111},
5241(1999).
\bibitem{15} F. Drolet and G. H. Fredrickson, \textit{Phys. Rev.
Lett.} \textbf{83}, 4317 (1999); \textit{Macromolecules}
\textbf{34}, 5317 (2001).
\bibitem{Den}P. Tarazona, \textit{Mol. Phys.} \textbf{52}, 81(1984).
\bibitem{sss}N. F. Carnahan and K. E. Starling, \textit{J. Chem. Phys.}  \textbf{51},
635(1969).
\bibitem{sma}The interaction of colloidal particles can be  adjusted by
chemically coating ligands onto the surfaces of them.
\bibitem{sda} T. Biben, J. P. Hansen, and J. L. Barrat, \textit{J. Chem. Phys.} \textbf{98},
7330(1993); C. I.  Addison, J. P. Hansen, and A. A. Louis,
\textit{ChemPhysChem.} \textbf{6}, 1760(2005).
\bibitem{sssso} M. Holgado   et al.,   \textit{Langmuir}
\textbf{15}, 4701(1999).
\bibitem{sssso1} M. Trau, D. A. Saville, and I.A. Aksay,
\textit{Science} \textbf{272}, 706(1996); M. Giersig and P.
Mulvaney, \textit{J. Phys. Chem.} \textbf{97}, 6334(1993);
\textit{Langmuir} \textbf{9}, 3408(1993).
\bibitem{ssb} N. V. Dziomkina and G. J. Vancso, \textit{Soft Matter}
\textbf{1}, 265(2005).
\bibitem{wang}  Y. Bohbot-Raviv and Z. G. Wang, \textit{Phys. Rev. Lett.} \textbf{85}, 3428(2000).
\bibitem{kang} We define $\varepsilon=
\max(h(x)) - \min(h(x))$, where the local brush height  $h(x)= 2
\sum_z \varphi(x,z)z/(\sum_z \varphi(x,z))$.
\bibitem{Fred} G. H. Fredrickson  et al., \textit{Macromolecules} \textbf{25}, 2882(1992).


\end{references}
\end{document}